\documentclass{article}

\usepackage{graphicx}
\usepackage{color}

\title{A methodological approach on the architectural development of integrated e-learning systems}
\author{Radu Gramatovici  and Ionut Tutu\\
Department of Computer Science, University of Bucharest}
\date{}

\begin{document}
\maketitle

\begin{abstract}
This study presents a methodological approach to the development of integrated e-learning systems that is used in the creation of educational content for standard Learning Management Systems. 
\end{abstract}

\section{Introduction}

The goal of this article is to describe a general methodology guided by well-established quality attributes used in the development of the architecture of an integrated e-learning system which will be further referred as \emph{IELS}.
Throughout the presentation we will refer to the reasons that led to various architectural decisions related to the division of the system into logical components and also to the interaction between the sub-systems involved in the development of the project.

The objective of IELS is to create an (online) integrated system that is to be used in the creation of educational content for standard \emph{LMSs} (Learning Management Systems). We believe that the main obstacle in the use of LMSs is the development of proper material on which learning sessions are based. IELS aims to fix this problem with intuitive and easy to use tools targeted towards the exact individuals that are going to guide the learners during the interactive lectures.

For start we set up five basic functionalities of IELS:
\begin{itemize}
\item a teacher will access the system to design a lesson,
\item he will be able to access the platform regardless on the operating system and the architecture of his machine,
\item the system will facilitate the development of attractive and intuitive educational content,
\item the system will be able to use existing content (from integrated repositories),
\item system requirements for both client and server will be minimized, with the possibility of extension for the server side.
\end{itemize}

Using standard technologies and accessible tools the teacher will be able to create, aggregate, reuse an publish educational content with ease. \\
The main actor is the content creator and one fundamental requirement of the system is that inexperienced users can be trained quickly referring to knowledge
acquired using other tools.

\section{Quality-driven Software Architecture Development}

The \emph{architecture} of a software system refers to the structural presentation of the system as a grouping of components that are connected through relations with specific properties. The software architecture also comprises the principles that conduct the design and development of the system.

\begin{figure}[h]
  \centering
  \includegraphics[scale=.75]{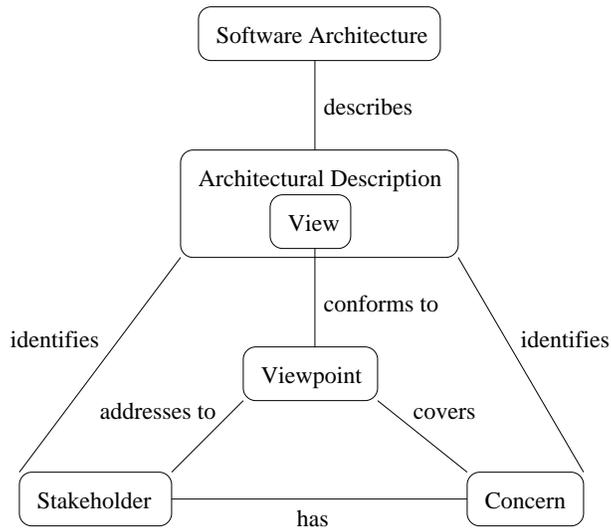}
  \caption{Conceptual framework}
  \label{fig:cf}
\end{figure}

The conceptual framework that lies at the foundations of the software architecture design is presented in figure~\ref{fig:cf}.
A starting point in developing a software system architecture lies in the definition of stakeholders and their concerns.

In developing a software architecture a \emph{stakeholder} is an entity (most often an individual, community or organization) which is involved in the development or use of the system and shows interest in the decisions that are made during the entire development phase. \\
During the development of IELS we have identified two classes of stakeholders:
\begin{itemize}
\item people who will interact with the deployed system: customers, owners, operators, system engineers,
\item people who are an active part in the development of the system: architects, designers, developers, distributors.
\end{itemize}

The \emph{concerns} of the stakeholders are generally focused on simplifying the development or use of the system. Starting from concerns one can identify issues or functional requirements of the system specific to each stakeholder.

The \emph{architectural description} includes those products which occur during the development of the system that reveal useful information about the architecture. These products can have various uses in:
\begin{itemize}
\item brief description of the system and its evolution,
\item communication between stakeholders,
\item assessment and comparison of architectures in a consistent way,
\item planning and execution of activities,
\item the degree of compatibility between the implementation and the architectural description.
\end{itemize}

In order to reduce the complexity of the architectural description one can identify \emph{views}, i.e. small collections of concerns, mostly related, of one or more stakeholders. From this perspective we can say that the architectural description is a consistent collection of views, each of them contributing with new information to the description of the architecture.

Each view is built on a set of resources and structuring rules well-defined by a \emph{viewpoint}. One can describe a viewpoint by specifying its name, the stakeholders and the concerns it addresses to and the modeling techniques used in building views.

Software architecture focuses on general decisions about the elements of the system and the interactions between them, thus making abstraction of more specific issues such as data structures or algorithms.
Based on the architecture one can derive a design plan that describes the hierarchical layout of the system and how elements of the system are integrated to form subsystems that meet requirements. Also the design plan can be used to sketch the development of the system, deal with possible answers to concerns of stakeholders and to set expectations for the clients.

\subsection{Quality Attributes}

From the perspective of software engineering, in the analysis of a system one can identify two types of requirements: functional and non-functional. \\
In general, functional requirements define the functionality of the system or what system should do. These requirements define the behavior of the system, referring to the transformations applied by components on some input data to get output data. \\
\emph{Non-functional requirements} are defined in terms of system properties or qualities. In contrast to functional requirements, non-functional requirements do not describe what the system does but focus on some characteristics of how the system operates. Examples of non-functional requirements include external interface requirements, design restrictions and \emph{quality attributes}.
Due to their nature, the task of assessing the response to a non-functional requirement is difficult and mostly subjective.

Depending on how their value is determined, quality attributes fall into two main categories:
\begin{itemize}
\item attributes that can be evaluated by analyzing the static structure of the system: scalability, extensibility, maintenance, testing effort, etc.
\item attributes that can be evaluated only while the system is running: performance, security, usability, etc.
\end{itemize}

In general, quality attributes are not unique to a development stage and should be considered throughout the entire development process, from design to implementation and deployment. 
Some quality attributes are considered critical so their value is assessed at the level of the architecture. Other attributes are too particular and can be evaluated only by analyzing the implementation.

In the development of IELS we have identified the following quality attributes:
\begin{itemize}
\item system qualities: availability, adaptability, maintenance, performance, security, creating logs, usability,
\item business qualities: time to market, integration, platform independence.
\end{itemize}

We will formally specify quality attributes through \emph{scenarios}. For each quality attribute a scenario should include:
\begin{description}
\item[source:] the software system or actor that produces a trigger,
\item[trigger:] the situation in which the quality attribute is evaluated,
\item[environment:] the state of the system,
\item[artifact:] the region of the system in which the assessment takes place,
\item[response:] the behavior of the system,
\item[assessment:] how the evaluation of the response is performed.
\end{description}

The \emph{performance} of a system refers to the number of tasks performed by the system in relation to used resources. The system performance is evaluated differently depending on context. In our project we will refer to four types of measurements:
\begin{itemize}
\item response time relative to a user's request,
\item rate of data processing,
\item utilization of computing resources,
\item volume of transmitted data.
\end{itemize}
The general scenario for performance is presented in table~\ref{table:gsp}.
\begin{table}[h]
  \begin{tabular}{r|p{.75\textwidth}}
    Source & independent sources \\
    Trigger & random or cyclic occurrences of events \\
    Environment & normal or overloaded operation \\
    Artifact & a processor, storage device or communication channel \\
    Response & change the level of service \\
    Assessment & latency, throughput, miss rate
  \end{tabular}
  \caption{General scenario for performance}
  \label{table:gsp}
\end{table}

The \emph{security} of a system refers to the capability of the system to ensure normal operation for the authorized user while resisting unauthorized accesses. In this context, unauthorized access may refer to:
\begin{itemize}
\item unauthorized attempts to access data,
\item unauthorized attempts to modify data,
\item actions targeted towards preventing the use of services by legitimate users.
\end{itemize}
The security of a system as a quality attribute can be divided into a number of sub-attributes:
\begin{itemize}
\item \emph{privacy}: data protection from unauthorized access,
\item \emph{integrity}: prevention of accidentally or maliciously alteration of data,
\item \emph{audit}: recording of data changes that occur in the system.
\end{itemize}
The general scenario for security is presented in table~\ref{table:gss}.
\begin{table}[h]
  \begin{tabular}{r|p{.75\textwidth}}
    Source & internal or external to the system, authorization, access \\
    Trigger & attempt to read, modify or delete data, to access system services or to alter the availability of services \\
    Environment & normal operation \\
    Artifact & system data, system services \\
    Response & providing data or services for legitimate users, prevention of unauthorized access \\
    Assessment & failure rate
  \end{tabular}
  \caption{General scenario for security}
  \label{table:gss}
\end{table}

The \emph{usability} of a system is directly reflected in the ease with which a user reaches its goal using data and services made available by the system. \\
Aspects related to usability are generally established in the prototype phase of the system and refer to:
\begin{itemize}
\item learning effort of a client to use the system,
\item efficiency of a client in using the system,
\item confidence of a client in using the system,
\item minimizing the negative impact of errors.
\end{itemize}
The general scenario for usability is presented in table~\ref{table:gsu}.
\begin{table}[h]
  \begin{tabular}{r|p{.75\textwidth}}
    Source & user \\
    Trigger & attempt to learn system features, use system efficiently and comfortable \\
    Environment & normal operation \\
    Artifact & system \\
    Response & providing system data and services \\
    Assessment & task time, user satisfaction, successful operations vs. failed operations
  \end{tabular}
  \caption{General scenario for usability}
  \label{table:gsu}
\end{table}

\subsection{Architectural tactics}

\emph{Architectural tactics} are fundamental design decisions that significantly contribute to the design and analysis of the system architecture. 
They are closely related to quality attributes and have a direct influence on the response to a single quality attribute. Architectural tactics are therefore sufficiently simple such that one could easily understand their properties and effects.

Most design decision have side effects, meaning that they affect other quality attributes besides those directly concerned. It follows that the use of an architectural tactic must be preceded by an estimation of both positive and negative impact of its implementation on other significant quality attributes.

We will further focus on architectural tactics aimed at the scenarios of three quality attributes described in the previous section: performance, security and usability.

\emph{Performance} targets various aspects of the system including computing time, response time, resource consumption, throughput and efficiency. We will simplify the analysis and concentrate on the response time relative to the messages or requests submitted by users. From this restricted point of view, performance related tactics can be classified as:
\begin{itemize}
\item resource management tactics,
\item resource arbitration tactics.
\end{itemize}

\emph{Resource management tactics} lead to improvement in the performance of the system by a better organization of the resources that have a major impact on the response time. \\
One such tactic is \emph{induced concurrency}. Under this tactic threads or distinct processes are associated with resources. Thus concurrent accesses on resources translates into concurrent executions of threads. In this way the waiting time can be reduced through techniques specific to concurrent processing, e.g. load balancing. \\
Another resource management tactic is \emph{targeted redundancy}. 
It is achieved by maintaining multiple copies, in various and usually distant regions of the system, of the data that is frequently used. The immediate consequence of the existence of multiple copies of the same resources is a weaker competition between the sub-systems that request access to those resources. In the case of raw data, using this tactic involves the existence of a data storage space with a cache and a cache manager. The cache aims to retain the data that is most often required, while the cache manager has to update the cache such that it reflects frequently requested data. It is also responsible with synchronizing the cache with the main storage space in situations when cached data gets changed.

\emph{Resource arbitration} is used to improve performance by scheduling requests to essential resources such as CPU or network. There are two well known methods of scheduling requests:
\begin{itemize}
\item \emph{FIFO} (First In First Out) tactics: all requests are treated equally in the order they are received,
\item tactics based on \emph{priorities}: the order the requests are processed is determined by (static or dynamic) priorities associated with each request.
\end{itemize}

The \emph{security} of the system deals with prevention of unauthorized access while providing services for legitimate users. Security related tactics can be tactics targeted towards resisting attacks, detecting attacks or recovering from attacks.

The first class of tactics offers several ways to protect the system during an attack. One protective measure is \emph{user authentication}, for example through IDs and passwords. In this case we have two possible tactics:
\begin{itemize}
\item ID / password: the ID and the password are chosen by the user,
\item Onetime password: the ID and the password are automatically generated by the system.
\end{itemize}

Another method of protection is the use of an integrated \emph{authorization system}. Under this tactic users' access to data and services is restricted depending on their privileges. These tactics are mostly used in combination with other user authentication tactics.

Other ways to ensure safety increase the \emph{resistance} of the system to attacks by using \emph{encrypted communication channels}. In such situations, specialized components are needed to efficiently encrypt / decrypt messages. \\
Resistance to attacks is strongly related to the \emph{detection} of attacks. Such tactics include limiting the attacker's possibilities, for example by disabling certain ports or through a firewall. Other methods consist of continuously monitoring
network traffic and compare it with patterns calculated from records of previous attacks.

\emph{Recovery tactics} have applicability in situations in which it is necessary to restore the system after a successful attack is recorded or as an attempt to minimize the damage produced by a successful attack. The basic restoration tactic consists of separating critical, security related data from ordinary data. This is achieved in IELS by separating the administrative data referring to the creators of educational content from actual data like the one used during learning sessions.

Tactics targeted towards \emph{usability} aim to develop a product with a low learning curve which users can use quickly and efficiently. Such tactics are divided into two categories:
\begin{itemize}
\item \emph{run-time tactics}: consist of interface design principles and are strongly influenced by customer feedback,
\item \emph{design-time tactics}: focused on the user interface design phase and on developers involved in this process.
\end{itemize}
According to these tactics, a number of interface design principles were considered in the development of IELS:
\begin{itemize}
\item early focus on users and activities through direct interactions between the interface design team and the target group,
\item testing the system since the first stages of development,
\item constant evaluation of the learning curve of the system,
\item applying an iterative design process via a cycle of prototyping, testing, analyzing, and refining.
\end{itemize}

\subsection{Architectural styles}

Patterns and architectural styles are ways of managing the complexity and size of the design of software architecture. Their use greatly reduces the process of designing the architecture to understanding, selection, combination or adjustment of well-established architectural patterns.

Because architectural patterns retain much of the complexity of the system, the results of their interaction are generally difficult to predict. In order to have a high degree of applicability architectural patterns are often specified generically. For this reason, during the design process, the patterns are altered with system-specific information.

An \emph{architectural pattern} shows an image of the system without being a complete architecture. More specifically an architectural pattern describes some essential elements of the software system architecture.

Similar to architectural tactics, an important characteristic of architectural patterns is that each pattern is related to a well-defined set of quality attributes. It follows that the choice of architectural patterns that provide the best response to requirements determined by quality attributes must be done in the early stages of design.

According to the notions presented in the previous sections, the description of a software architecture includes:
\begin{itemize}
\item \emph{components} responsible with computations or storing data: objects, filters, databases, etc.
\item \emph{connectors} that ensure interaction between components: procedure calls, communication channels, events, etc.
\item \emph{attributes} that provide the useful information in analysis and design of the architecture: signatures, pre-conditions, post-conditions, etc.
\end{itemize}

Components are the basic elements used in the development of the architecture. Most of the times they are active entities of the system, with well-defined computational properties.
Communication of a component with the environment is done through ports or interfaces. A port may be a user interface, a shared variable, a reference to a procedure of another component, a collection of events that can be triggered, etc.

Each component defines one or more interfaces that specify its communication capabilities. Each interface can connect to multiple interfaces of other components. The analysis of the communication between components must consider a number of attributes like the direction of data flow, the existence of a buffer and its capacity, the supported communication protocols, etc.

An \emph{architectural style} defines a family of architectures that have a common architectural description, topology and other semantic constraints.

In this context, the \emph{topology} or configuration of an architectural style is a graph in which vertices and edges are the core elements of the architectural description: components and connectors. Thus the analysis of components and connectors extends to the analysis of topologies. Frequent tests are related to the overall performance of an architecture, concurrent and distributed computing properties, reliability, etc.

In general the architecture of a system is obtained by combining several architectural styles. The architecture considered for IELS falls into this category although the predominant model is the \emph{RIA} (Rich Internet Application) client-server model.

\emph{RIAs} are web applications with most of the features of regular desktop applications but generally run in a web browser or a virtual machine. Among the most popular RIA technologies we mention Flex, Ajax, JavaFX and Silverlight. Applications developed using these technologies share a similar architecture based on a client module and a distinct level of services.

RIA is more suited to a client-server development than to a traditional web development where the state of the system is part of the session. In a client-server approach the client has its own state, it knows the data it needs and also the types of the data received from the server.

Among the immediate benefits of this approach we mention a well organized service level, simple requests to the server and a reduced computation demand on the server by running some calculations on the client side.

The global architecture of a \emph{client-server application} includes a series of sub-architectures. The system can be briefly described as the communication between the client and the server using a level of services. Each component of the global architecture has its own local architecture.

\emph{MVC} (Model-View-Controller) is an architectural pattern in which the system is divided into a series of sub-systems responsible for data access, business logic and user interface.
\begin{figure}[h]
  \centering
  \includegraphics[scale=.75]{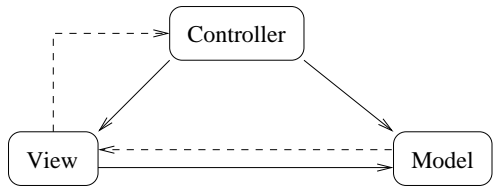}
  \caption{The Model-View-Controller architectural pattern}
\end{figure}

The name of the pattern is given by the roles of its components:
\begin{description}
\item[The model] is a convenient representation of data on which the application runs. When the state of a model changes the views associated with the model are automatically informed to update the interface. Generally it is assumed that the level of data access is integrated within the model.
\item[The view] is generally an element specific to the user interface that displays a particular model. More insights on the model can be obtained using multiple views. Views are also responsible for interacting with the user.
\item[The controller] provides communication between views and models. A control module receives input and generates a response after querying a model.
\end{description}

In a RIA application one can identify two levels that implement the MVC architectural pattern:
\begin{itemize}
\item an implementation at the client level,
\item an implementation at the service level.
\end{itemize}
Although both levels display features specific to MVCs, a complete functionality is rarely implemented.

The \emph{client MVC} manages the interaction between user and user interface (by invoking commands, loading data, updating the interface etc.). The main objectives of a client MVC are maintaining the state of the application, mediating requests to the server and displaying control data.

The task of the \emph{server MVC} is to manage the requests received from the client.
For this it processes the requests and triggers actions on the server. 
These actions may include:
\begin{itemize}
\item saving information in the database,
\item updating information,
\item returning requested information,
\item analytical calculations.
\end{itemize}
The major difference compared to the corresponding client MVC is that in this situation there is no user interface. Instead, views consist in the format of the data returned to the client application.

Common technologies appropriate to RIA development can be:
\begin{itemize}
\item Java, PHP, Rails, .NET for the service level,
\item Flex, AJAX, JavaFX for the client.
\end{itemize}








\section{Acknowledgements}

This work was partially supported by the grant of the Minsitry of Education, Research, Youth and Sports, no. 136/2007, through the program "Innovation".

\end{document}